\documentclass[aps,twocolumn,prl,showpacs]{revtex4}
\usepackage{amsfonts}

\usepackage{graphicx}
\usepackage{amsmath,amssymb,amsthm,bbm,latexsym}
\usepackage{amsthm}
\usepackage{amsbsy}
\usepackage{epsfig}
\usepackage{graphicx}
\usepackage{float}
\usepackage{dcolumn}
\usepackage{amsmath}

\begin{document}

\title{Violation of monogamy inequality for higher-dimensional
objects}

\begin{abstract}
Bipartite quantum entanglement for qutrits and higher dimensional
objects is considered. We analyze the possibility of violation of
monogamy inequality, introduced by Coffman, Kundu, and Wootters, for
some systems composed of such objects. An explicit counterexample
with a three-qutrit totally antisymmetric state is presented. Since
three-tangle has been confirmed to be a natural measure of
entanglement for qubit systems, our result shows that the
three-tangle is no longer a legitimate measure of entanglement for
states with three qutrits or higher dimensional objects.
\end{abstract}

\author{Yong-Cheng Ou }
\email{ouyongcheng@163.com}
 \affiliation{Institute of Physics,
Chinese Academy of Sciences, Beijing 100080, P.R.China}
 \maketitle

Quantification of quantum entanglement plays an important role not
only in quantum information processing and quantum
computation\cite{nielsen} but also in describing quantum phase
transition in various interacting quantum many-body
systems\cite{osterloh}. In the last ten years a number of
entanglement measures for qubit systems have been studied
extensively, in which the well known one with an elegant formula is
concurrence derived analytically by Wootters\cite{wootters}, and the
entanglement of formation(EOF)\cite{bennett, hill} is a
monotonically increasing function of the concurrence. However, at
least so far it is believed that there exists a drawback that they
are confined into the qubit systems since the used spin-flip is only
applicable to qubits\cite{werner}, Because of which generally only a
lower bound of concurrence can be achieved for states composed of
qutrits or higher dimensional objects. The seminal paper by Coffman,
Kundu, and Wootters\cite{coffman} provided a basis for the
quantification of three-party entanglement by introducing the so
called residual tangle, and a general monogamy inequality in the
case of $n$ qubits has been proved\cite{osborne}.

Since the monogamy inequality has been established, whether it can
be generalized to qutrits or higher dimensional objects remains
still open. In this Brief Report, we will firstly show that the
monogamy inequality can be violated for some quantum composed of
qutrits or high dimensional objects, and then offer an explicit
example of an antisymmetric state to show this violation. Therefore
the main idea here is to show that the monogamy inequality
characterized by the concurrence can not be generalized to quantum
state apart from qubits. This result gives a caveat when we are
studying genuine multipartite entanglement for such states where the
residual entanglement or 3-tangle is defined.

For completeness we recall the original monogamy inequality.
Consider a triple of spin-1/2 particles $A$, $B$, and $C$, and its
density matrix is denoted by $\rho_{ABC}$, the distribution of
entanglement among them is constrained by the following inequality
\begin{equation}  \label{a}
C_{AB}^{2}+C_{AC}^{2}\leq C_{A(BC)}^{2},
\end{equation}
where $C_{AB}$ and $C_{AC}$  are the concurrences of the state
$\rho_{ABC}$ with traces taken over the particles  $C$, $B$.
$C_{A(BC)}$ is the concurrence of $\rho_{A(BC)}$ with the particles
$B$ and $C$ regarded as a single object. In this case the particle
$A$ can be viewed as a focus such that the three-tangle can be
defined as
\begin{equation}\label{2}
\tau_{ABC}=C_{A(BC)}^{2}-C_{AB}^{2}-C_{AC}^{2}.
\end{equation}
which is independent on the choice of the focus mainly because it is
invariant under the permutations of the particles. The three-tangle
has found wide applications in the research of genuine
multi-particle entanglement\cite{cai} since it satisfies necessary
conditions that a natural entanglement measure for pure state must
require of\cite{vedral}. (1) $\tau_{ABC}$ is invariant under local
unitary operations. (2) $\tau_{ABC}\geq 0$ for all pure states. (3)
$\tau_{ABC}$ is an entanglement monotone\cite{due}, i.e., it does
not increase on average under local quantum operations assisted with
classical communication.

Now one may naturally ask what will happen as a general $M\times N
\times Q$ system is concerned. It's in our great interest to know
whether Eq.(1) remains valid since it determines justification of
$\tau_{ABC}$ as an natural entanglement measure. In order to get the
answer,
we review some progresses on the quantification of entanglement of a bipartite $%
M\times N$ system. Actually only a few special classes of high
dimensional states can give a closed-form expression of
EOF\cite{vidal1}\cite{fei}. Through analytical or numerical
approaches generally only a lower bound of concurrence or EOF can be
obtained\cite{mintert,chen,song}. The unavailability of exact
entanglement makes us suspect validity of corresponding monogamy
inequality in Eq.(1) for higher dimensional tripartite systems.

In what follows we show there exists a possibility that Eq.(1) does
not necessarily hold always for qutrits or higher dimensional
objects. For a pure bipartite $M\times N$
system the squared concurrence can be expressed as $%
C^{2}=\sum_{\alpha =1}^{M(M-1)}\sum_{\beta =1}^{N(N-1)}|C_{\alpha
\beta }|^{2}$\cite{sj}, where $C_{\alpha \beta }=\langle \psi
|L_{\alpha }\otimes L_{\beta
}|\psi ^{\ast }\rangle $, and $L_{\alpha }$, $L_{\beta }$ are generators of $%
SO(M)$ and $SO(N)$, respectively. For mixed state $\rho $ of such system the
squared concurrence is given as the convex roof
\begin{equation}
\begin{array}{cc}
C^{2}(\rho )=\mathtt{inf}\sum\limits_{i}p_{i}C^{2}(|\Phi _{i}\rangle ), &
\rho =\sum\limits_{i}p_{i}|\Phi _{i}\rangle \langle \Phi _{i}|,%
\end{array}
\label{b}
\end{equation}%
where $p_{i}\geq 0$ and $\rho $ consists of all possible
decompositions into pure states $|\Phi _{i}\rangle $. According to
\cite{mintert,song}  we can obtain a lower bound
$\mathbb{C}^{2}(\rho)$ for the mixed state as
\begin{equation}
\mathbb{C}^{2}(\rho )\equiv \left( \lambda _{1}-\sum_{i>1}\lambda
_{i}\right) ^{2} \leq \mathcal{C}^{2}(\rho ),  \label{c}
\end{equation}%
where $\lambda _{i}$ are the singular values of $\sum_{\alpha
=1}^{M(M-1)}\sum_{\beta =1}^{N(N-1)}z_{\alpha \beta }A_{\alpha \beta
}$ in decreasing order. For details of terms $z_{\alpha \beta }$ and
$A_{\alpha \beta }$, see\cite{mintert, song}. Note that the choice
of the phase of $z_{\alpha \beta }$ is important for the tightness
of the bound and in general it need complicated numerical
optimization procedure for the bound. For the simplest  $2\times 2$
systems, Eq.(4) becomes equality and analytical. Moreover, the lower
bound can also be written as an analytical expression
\begin{equation}\label{6}
\mathbb{C}^{2}(\rho )=\sum_{j>
i}\sum_{i=1}^{M-1}\mathcal{C}^{2}_{ij}(\rho),
\end{equation}
for the more complicated $2\times M$ systems\cite{ger}, where
$\mathcal{C}_{ij}(\rho)=\mathrm{max}(0,\lambda_{1}^{ij}-\lambda_{2}^{ij}-\lambda_{3}^{ij}-\lambda_{4}^{ij})$
and $\lambda^{ij}$ are the square roots of the four largest
eigenvalues of the matrix
$\rho^{1/2}S^{ij}\rho^{*}S^{ij}\rho^{1/2}$\cite{ger}. It does not
need numerical optimization for the bound of concurrence of the
$2\times M$ systems.

Let us consider the tripartite $2\times M\times N$ system $ABC$.
Choose the particle $A$ with two dimension as a focus, it then
follows from Eq.(4) that the lower bound of the squared concurrences
in $AB$ and $AC$ satisfies
\begin{equation}  \label{d}
\begin{array}{cc}
\mathbb{C}^{2}_{AB}\leq C^{2}_{AB}, & \mathbb{C}^{2}_{AC} \leq C^{2}_{AC}.%
\end{array}%
\end{equation}
Combining the two inequalities in Eq.(6) gives
\begin{equation}  \label{f}
\mathbb{C}^{2}_{AB}+\mathbb{C}^{2}_{AC}\leq C^{2}_{AB}+C^{2}_{AC}.
\end{equation}
On the other hand, the authors in \cite{song} have proven that the
sum of the lower bound in  $\mathcal{C}^{2}_{AB}$ and
$\mathcal{C}^{2}_{AC}$ is not greater than the squared concurrence
between $A$ and $BC$
\begin{equation}  \label{e}
\mathbb{C}^{2}_{AB}+\mathbb{C}^{2}_{AC}\leq C^{2}_{A(BC)}.
\end{equation}
From the observation of Eqs.(7) and (8), it is possible for us to
find some states making Eq.(8) be equality but Eq.(7) be strict
inequality, thus resulting in the violation of the monogamy
inequality
\begin{equation}  \label{e2}
\mathcal{C}^{2}_{AB}+\mathcal{C}^{2}_{AC}\geq C^{2}_{A(BC)}.
\end{equation}
Note that Eq.(7) can only be an equality for the $2\times 2 \times
2$ systems, making the monogamy inequality in Eq.(1) hold for each
state in such system. In practice it is a formidable task to find
the state satisfying Eq.(9) because of the requirement of
complicated convex roof for the concurrence of higher dimensional
mixed state. Fortunately we find a state in the following,
permitting us to easily calculate the concurrence of the mixed
state.

Finally we present the explicit example that the monogamy inequality
dose not work. Consider the pure totally antisymmetric state on a
three-qutrit system $ABC$
\begin{equation}  \label{f}
|\Psi\rangle=\frac{1}{\sqrt{6}}\left(|123\rangle-|132\rangle+|231%
\rangle-|213\rangle+|312\rangle-|321\rangle\right).
\end{equation}
It is obvious that antisymmetric subspace $V\in H_{A}\otimes H_{B}$, $%
H_{A}\otimes H_{C}$, and $H_{B}\otimes H_{C}$, is spanned by the vectors
\begin{equation}  \label{g}
\begin{array}{c}
|x\rangle_{ij}\equiv \frac{1}{\sqrt{2}}\left(|23\rangle-|32\rangle\right), \\
\\
|y\rangle_{ij}\equiv \frac{1}{\sqrt{2}}\left(|31\rangle-|13\rangle\right), \\
\\
|z\rangle_{ij}\equiv \frac{1}{\sqrt{2}}\left(|12\rangle-|21\rangle\right),%
\end{array}%
\end{equation}
where $\{ij\}\in \{AB, AC, BC\}$. If $A$ is chosen as a focus, then
\begin{equation}  \label{h}
\begin{array}{c}
\rho_{AB}=\frac{1}{3}\left(|x\rangle_{AB}\langle x|+|y\rangle_{AB}\langle
y|+|z\rangle_{AB}\langle z|\right), \\
\\
\rho_{AC}=\frac{1}{3}\left(|x\rangle_{AC}\langle x|+|y\rangle_{AC}\langle
y|+|z\rangle_{AC}\langle z|\right), \\
\\
\rho_{A(BC)}=\frac{1}{3}\left(|x\rangle_{BC}\langle x|+|y\rangle_{BC}\langle
y|+|z\rangle_{BC}\langle z|\right).%
\end{array}%
\end{equation}
Since $\rho _{A(BC)}$ is pure, it is easy to check
$C^{2}_{A(BC)}=4/3$. For mixed states $\rho_{AB}$ and $\rho_{BC}$ we
have to make an infimum for their concurrences. Generally, it is
difficult, however, the system (10) is a
special case. For arbitrary pure states $|\Phi\rangle_{AB}=c_{1}|x%
\rangle+c_{2}|y\rangle + c_{3}|z\rangle$ with $%
|c_{1}|^{2}+|c_{2}|^{2}+|c_{3}|^{2}=1$, their reduced density matrix $%
\rho_{A}\equiv \mathtt{Tr}_{B}|\Phi\rangle_{AB}\langle\Phi|$ has the same
spectrum $\{1/2, 1/2, 0\}$\cite{vidal}, implying any two antisymmetric
states $|\Phi\rangle_{AB}$ can be transformed into each other by local
unitary transformations. As a result, $C^{2}(|\Phi\rangle_{AB})=1$. While $%
\rho _{AB}$ in Eq.(9) can be decomposed into
\begin{equation}  \label{j}
\rho_{AB}=\sum_{i}p_{i}|\Phi_{i}\rangle_{AB}\langle\Phi_{i}|.
\end{equation}
Why the system (10) is special lies in
$C^{2}(|\Phi_{i}\rangle_{AB})=1$ for each $|\Phi_{i}\rangle_{AB}$
such that $C^{2}_{AB}=\sum_{i}p_{i}=1$. Analogously,
$C^{2}_{AC}=\sum_{i}p_{i}=1$. Therefore we obtain
\begin{equation}  \label{i}
C_{AB}^{2}+C_{AC}^{2}=2\geq \frac{4}{3}=C_{A(BC)}^{2},
\end{equation}
which is not superseded by the monogamy inequality in Eq.(1). In the
similar way, it is confirmed that $C_{BA}^{2}+C_{BC}^{2}\geq
C_{B(AC)}^{2}$ and $C_{CA}^{2}+C_{CB}^{2}\geq C_{C(AB)}^{2}$ also
violate the corresponding monogamy inequalities. Perhaps this
violation is not a paradox by considering that the EOF can not yet
satisfy the monogamy inequality due to the concave log
function\cite{coffman}.

Summarizing, we have shown that the monogamy inequality in qubit
systems can not be generalized to higher dimensional objects such
that a caveat is provided when the three-tangle is defined since it
may exhibit a negative value for some state. However, in general the
monogamy inequality in Eq.(1) also works, for example, for a state
$|\Psi\rangle_{ABC}=\frac{1}{\sqrt{3}}(|111\rangle+|222\rangle+|333\rangle)$,
we obtain $C_{AB}^{2}+C_{AC}^{2}=0\leq 4/3=C_{A(BC)}^{2}$,
satisfying the monogamy inequality. Consequently, the conditions of
whether the monogamy inequality for
higher dimensional objects is violated or not is still open. As stated in%
\cite{osborne}, the constrains by the monogamy inequality in Eq.(1)
on the entanglement shared by parties lie at the heart of the
success of many information-theoretic protocols, correspondingly
the impacts on such protocols imposed by this violation of monogamy
inequality deserve further investigations.

The author thanks Heng Fan for many valuable discussions. This
project was granted financial support from China Postdoctoral
Science Foundation.

\end{document}